\documentclass[12pt]{iopart}
\pdfoutput=1

\expandafter\let\csname equation*\endcsname\relax
\expandafter\let\csname endequation*\endcsname\relax

\usepackage{color}
\usepackage{graphicx}
\usepackage{amsmath,amssymb,wasysym}
\usepackage[sort&compress,numbers]{natbib}
\usepackage{mathrsfs}

\begin{document}
\date{\today}

\title[Thermal FF regime in a LJJ]{Thermal flux-flow regime in long Josephson tunnel junctions}

\author{Claudio Guarcello$^1$, Paolo Solinas$^2$, Francesco Giazotto$^3$, Alessandro Braggio$^3$}

\address{$^1$ Centro de Fisica de Materiales (CFM-MPC), Centro Mixto CSIC-UPV/EHU, 20018 Donostia-San Sebastian, Basque Country, Spain}
\address{$^2$ SPIN-CNR, Via Dodecaneso 33, 16146 Genova, Italy}
\address{$^3$ NEST, Istituto Nanoscienze-CNR and Scuola Normale Superiore, Piazza San Silvestro 12, I-56127 Pisa, Italy}
\ead{claudio.guarcello@ehu.eus}

\begin{abstract}
We study thermal transport induced by soliton dynamics in a long Josephson tunnel junction operating in the flux-flow regime. A thermal bias across the junction is established by imposing the superconducting electrodes to reside at different temperatures, when solitons flow along the junction. Here, we consider the effect of both a bias current and an external magnetic field on the thermal evolution of the device. In the flux-flow regime, a chain of magnetically-excited solitons rapidly moves along the junction driven by the bias current. We explore the range of bias current triggering the flux-flow regime at fixed values of magnetic field, and the stationary temperature distribution in this operation mode. We evidence a steady multi-peaked temperature profile which reflects on the average soliton distribution along the junction. Finally, we analyse also how the friction affecting the soliton dynamics influences the thermal evolution of the system.

\end{abstract}


\maketitle

\section{Introduction}
\label{Intro}\vskip-0.2cm

The possibility of mastering the local temperature of a long Josephson junction (LJJ) by acting on solitonic excitations is scientifically intriguing and it has an applicative potential in fast heat mastering. Since a soliton is a magnetic flux quantum surrounded by a loop of dissipationless supercurrents, how can it affect thermal transport through the system? In this regard, it was recently discussed theoretically~\cite{Gua18} that in a long Josephson tunnel junction a steady localized 2$\pi$-twist of the phase, that is a soliton~\cite{Bar71,Par78,McL78,Ust98,Mal14,Maz14}, is able to locally affect the quasiparticle heat-current flowing through a junction formed by superconducting electrodes residing at different temperatures. In this case, the emerging temperature modulation in correspondence of a soliton is not ascribed to a direct Cooper pairs contribution, but it is a local phase-dependent modulation of heat carried by quasiparticles flowing from the hot to the cold electrode. In fact, after the earlier theoretical prediction that heat transport can depend on the Josephson phase difference~\cite{Mak65,Gut97,Gut98,Zha03,Zha04}, this phenomenon was recently confirmed experimentally in several temperature-biased Josephson devices~\cite{ForGia17}. This phenomenon gives, for instance, the capability to control the temperature of the system via an external magnetic field, as it was demonstrated, both theoretically and experimentally, in heat interferometers~\cite{Gia12,GiaMar12,ForBla16,Gua17} and quantum diffractors~\cite{Gia13,Mar14,Gua16} of heat currents. These examples fall within the so-called phase-dependent caloritronics~\cite{MarSol14,ForGia17} from which many temperature-based novel devices were recently conceived~\cite{For14,ForMar15,For16,Sol16,Hof16,Sot17,Hwa18,GuaSol18,Kam19,Gua19}. In a long Josephson tunnel junction, i.e., a junction in which one dimension is longer than the Josephson penetration length~\cite{Bar82}, the externally applied magnetic field can penetrate the junction in the form of fluxons. This kind of excitations can be controlled and handled in different ways, for instance they can be moved by a bias current, or created by a magnetic field or a dissipative hotspot~\cite{Mal94,Dod97}, pinned by inhomogeneities~\cite{Ust91,Feh92}, and also manipulated through shape engineering~\cite{Car02,Gul07,Cas18,Marin18}. Additionally, it was recently understood that solitons can induce thermal effects in a temperature-biased junction, so that applications as thermal router~\cite{Gua18,GuaSolBra18} and heat oscillator~\cite{GuaSolBraGia18} have been suggested.

In this paper we give a further insight in the research field of phase-dependent caloritronics based on LJJs. In fact, here we explore the effects of both an external magnetic field and a bias current on thermal transport through a temperature-biased LJJ. In particular, we investigate the so-called \emph{flux-flow} regime~\cite{Ern80,Par93,Gol01}, that is the case in which solitons in the form of fluxons are continuously magnetically excited from one edge of the junction and then forced to shift towards the opposite junction edge under the action of a bias current. In this operation mode, we observe some peculiar thermal effects depending on the dynamical state of solitons excited along the system. In particular, despite solitons move very rapidly, we observe temperature rises inhomogeneously in specific points of the system. We study also how friction affecting the phase dynamics influences the stationary temperature distribution.

The number of applications in different fields based on LJJs is still nowadays growing~\cite{Ooi07,Lik12,Sol14,Fed14,Sol15,Hil18,Osb18,Fra19}, not to mention that ones in which LJJs are used in both flux-flow regime and oscillator~\cite{Pan08,Mat11,Rev12,Pan15,Gul17}. How a homogeneous temperature gradient applied along a LJJ (namely, from one edge of the junction to the other) affects soliton dynamics was earlier studied both theoretically and experimentally~\cite{Log94,Gol95,Kra97}, but the soliton-sustained thermal transport as a temperature gradient is imposed across the system (namely, as the electrodes forming the junction reside at different temperatures) was exclusively addressed recently~\cite{Gua18,GuaSolBra18,GuaSolBraGia18}. In this regard, both the heat oscillator application~\cite{GuaSolBraGia18} and the thermal router~\cite{Gua18} readily lend themself to a further advance in the flux-flow operation mode.


This paper is organised as follows. In the next section both the sine-Gordon equation and thermal model are presented. 
In section~\ref{Results} the theoretical results, including the temperature dynamics as a function of the bias current, the external magnetic field, and the damping parameter, are shown and analysed. 
Finally, in section~\ref{Conclusions} the conclusions are drawn.

\begin{figure}[t!!]
\begin{center}
\includegraphics[width=0.7\columnwidth]{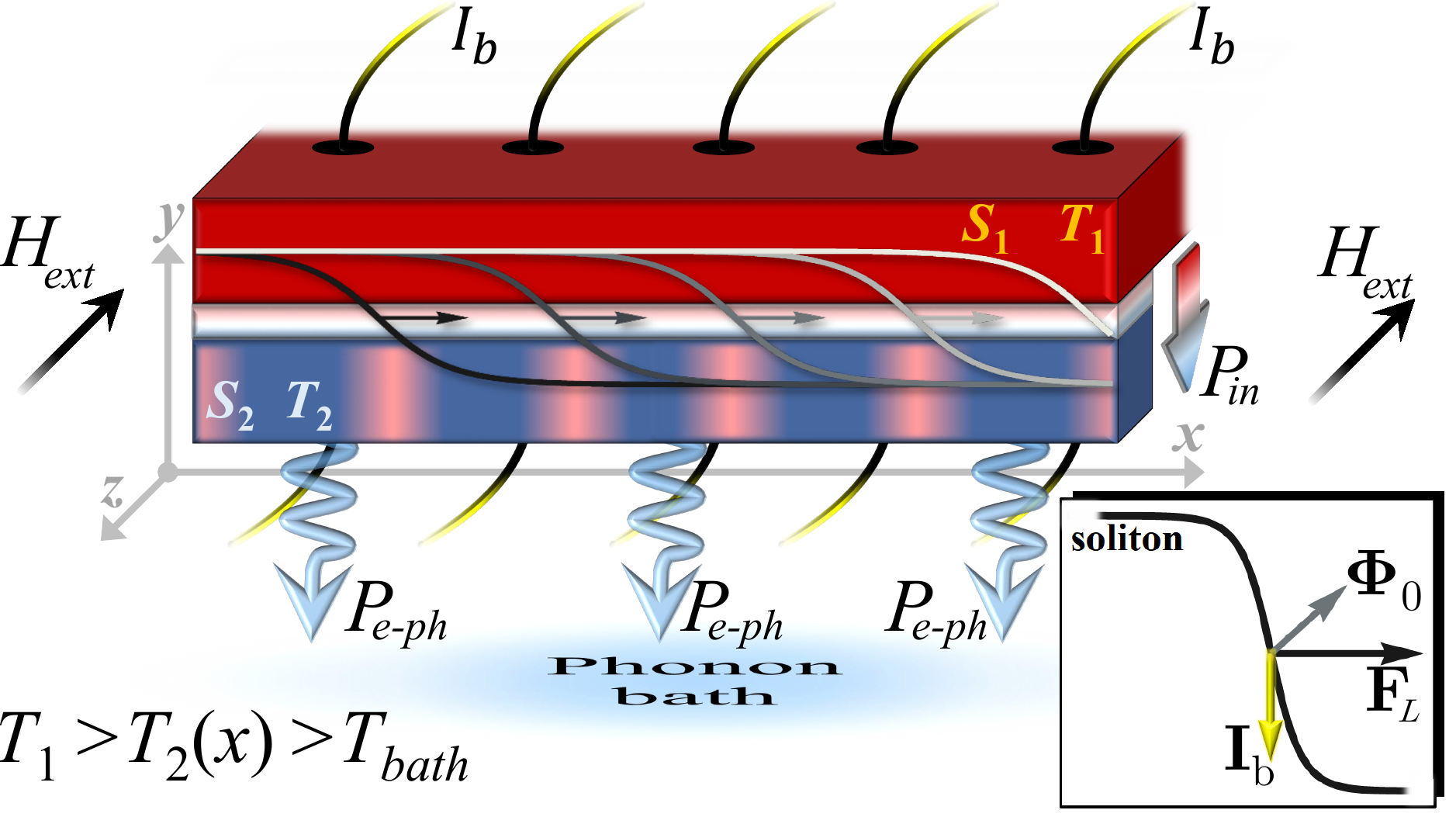}
\caption{A tunnel LJJ driven by both an external in-plane magnetic field, $H_{\text{ext}}(t)$, and a homogeneously distributed bias current, $I_b$. The temperature $T_i$ of each electrode $S_i$ is also indicated. A chain of solitons drifting along the junction under the action of $I_b$ is depicted. The incoming, i.e., $P_{\text{in}}\left ( T_1,T_2,\varphi,V \right )$, and outgoing, i.e., $P_{e\text{-ph}}\left ( T_2,T_{\text{bath}}\right )$, thermal powers in $S_2$ are also represented, for $T_1>T_2(x)>T_{\text{bath}}$. In the inset: bias current-induced Lorentz force, $\mathbf{F}_L\propto\mathbf{I}_\text{b}\times\mathbf{\Phi}_0$, acting on a soliton [where the direction of $\mathbf{\Phi}_0$ depends on the polarity, $\sigma$, of the soliton, see Eq.~\eqref{SGkink}].} 
\label{Figure01}
\end{center}
\end{figure}

\section{The Model}
\label{Model}\vskip-0.2cm 
%

The system is driven by both an external magnetic field, $H_{\text{ext}}(t)$, applied in both sides of the device and a homogeneous bias current, $I_{\text{b}}$, flowing through the junction, see Fig.~\ref{Figure01}. The behavior of a long and narrow Josephson tunnel junction depends on the dynamics of the Josephson phase $\varphi$, which can be described by the perturbed sine-Gordon (SG) equation~\cite{Bar82} 
\begin{equation}
\frac{\partial^2 \varphi\big(\widetilde{x},\widetilde{t}\,\big) }{\partial {\widetilde{x}}^2} -\frac{\partial^2 \varphi\big(\widetilde{x},\widetilde{t}\,\big) }{\partial {\widetilde{t}}^{2}}- \sin \left [\varphi\big ( \widetilde{x},\widetilde{t} \,\big ) \right ] = \alpha\frac{\partial \varphi\big(\widetilde{x},\widetilde{t}\,\big) }{\partial \widetilde{t}}+\widetilde{I}_b.\label{SGeq}
\end{equation}
In this formulation, we used normalized space and time variables, $\widetilde{x}=x/\lambda_{_{\text{J}}}$ and $\widetilde{t}=\omega_pt$, respectively, where $\lambda_{_{\text{J}}}=\sqrt{\frac{\Phi_0}{2\pi \mu_0}\frac{1}{t_d J_c}}$ is the Josephson penetration depth and $\omega_p=\sqrt{\frac{2\pi}{\Phi_0}\frac{J_c}{C}}$ is the Josephson plasma frequency. 
Here, we introduced the critical current area density $J_c=I_c/(L\times W)$ (where $L$ and $W$ are the length and the width of the junction, respectively) and the effective magnetic thickness~\cite{Gia13,Mar14} $t_d=\lambda_{L,1}\tanh\left ( d_1/2\lambda_{L,1} \right )+\lambda_{L,2}\tanh\left ( d_2/2\lambda_{L,2} \right )+d$ (where $\lambda_{L,i}$ and $d_i$ are the London penetration depth and the thickness of the electrode $S_i$, respectively, and $d$ is the insulating layer thickness). Moreover, $\mu_0$ is the vacuum permeability and $\Phi_0= h/2e\simeq2\times10^{-15}\; \textup{Wb}$ is the magnetic flux quantum, with $e$ and $h$ being the electron charge and the Planck constant, respectively.
We point out that we make use in this work of a notation in which a tilde over a letter labels a dimensionless normalized quantity, so that, for instance, the term $\widetilde{I}_b=I_b/I_c$ in Eq.~\eqref{SGeq} indicates the normalized bias current flowing through the junction.
The friction in the phase dynamics is accounted by the damping parameter $\alpha=1/(\omega_pR_aC)$, with $R_a$ and $C$ being the normal-state resistance per area and the specific capacitance of the junction, respectively~\cite{Tin04}. For simplicity, in the SG model we neglect the surface losses in the electrodes~\footnote{For moderate fluxon velocities, the effective damping term including both the quasiparticle tunneling and the surface current losses can be accounted by $\alpha_{eff}=\alpha+\beta/3$~\cite{McL78}, with $\beta$ being the parameter quantifying the surface losses of the junction.}.

The Josephson penetration depth $\lambda_{_{\text{J}}}$ represents the main length-scale in our system, so that the junction is called ``long'' when its length and width, in units of $\lambda_{_{\text{J}}}$, read $\widetilde{L}=L/ \lambda_{_{\text{J}}}\gg1$ and $\widetilde{W}=W/ \lambda_{_{\text{J}}}\ll1$, respectively. Moreover, $\lambda_{_{\text{J}}}$ roughly indicates also the width of a \emph{soliton}~\cite{Par93,Ust98}. This is a 2$\pi$-twists of the phase that in the LJJ framework has a clear physical meaning, since it carries a quantum of magnetic flux, induced by a supercurrent loop surrounding it, with the local magnetic field perpendicularly oriented with respect to the junction length. Thus, solitons in the context of LJJs are also indicated as \emph{fluxons} or \emph{Josephson vortices}. In the unperturbed case, i.e., Eq.~\eqref{SGeq} with no drive and dissipation, a moving soliton has the simple analytical expression~\cite{Bar82}
\begin{equation}
\varphi\big(\widetilde{x}-\widetilde{u}\,\widetilde{t}\,\big)=4\arctan \left [ \exp \left ( \sigma \frac{\widetilde{x}-\widetilde{x}_0-\widetilde{u}\,\widetilde{t} }{\sqrt{1-\widetilde{u}^2}} \right ) \right ],
\label{SGkink}
\end{equation}
where $\sigma=\pm1$ is the polarity (so that the $+$ sign indicates a soliton and the $-$ sign indicates an antisoliton) and $\widetilde{u}$ is the speed of the soliton, given in units of the Swihart’s velocity $\bar{c}=\lambda_{_{\text{J}}}\omega_p$~\cite{Bar82}. The latter is the phase velocity of electromagnetic waves propagating in the junction, and can approach the values $\bar{c}\sim10^6-10^7\;\text{m}/\text{s}$ in high-quality tunnel LJJs. 
The velocity-dependent factor in Eq.~\eqref{SGkink} represents the relativistic contraction of the soliton when its velocity approaches the maximum speed~\cite{McL78}. This is the consequence of Lorentz invariance of the unperturbed SG equation~\cite{Bar82}.


An external magnetic field, $H_{\text{ext}}$, affects the phase dynamics, since it is accounted by boundary conditions of Eq.~\eqref{SGeq},
\begin{equation}
\frac{d\varphi(0,t) }{d\widetilde{x}} = \frac{d\varphi(\widetilde{L},t) }{d\widetilde{x}}=2\frac{H_{\text{ext}}}{H_{c,1}}= \widetilde{H}.
\label{bcSGeq}
\end{equation}
The coefficient $H_{c,1}=\frac{\Phi_0}{\pi \mu_0 t_d\lambda_{_{\text{J}}}}$ is called the first critical field of a LJJ~\cite{Gol01}. So, for magnetic fields $H_{\text{ext}}$ exceeding this critical value, that means for $\widetilde{H}\geq \widetilde{H}^{\text{thr}}$ with $\widetilde{H}^{\text{thr}}=2$, solitons penetrate the junction also in the absence of an applied bias current.

When a bias current is flowing through the system the situation changes, since it exerts a Lorentz force, $\mathbf{F}_L\propto\mathbf{I}_\text{b}\times\mathbf{\Phi}_0$, on a soliton, see the inset of Fig.~\ref{Figure01}, with the direction of $\mathbf{\Phi}_0$ depending on the polarity of the soliton, see Eq.~\eqref{SGkink}. Thus, in the presence of an external bias current the soliton is forced to shift along the junction. In the case of several solitons excited in the system, the chain of solitons moves against the damping forces under the action of the Lorentz force exerted by the bias current. Reaching an edge of the junction, the fluxon leaves the system, while a new fluxon enters the junction from the opposite edge. This operation mode is called \emph{flux-flow} regime~\cite{Ern80,Par93,Gol01}. In this state, the magnetic flux penetrates effectively the junction in the form of fluxons.

In Ref.~\cite{Gua18}, it was demonstrated that the phase distribution along a LJJ affects thermal transport through the system, when a temperature bias is imposed. 
In this work we investigate the time evolution of the temperature $T_2$ of the floating electrode $S_2$ by changing both the bias current flowing through the system and the applied magnetic field. 
Specifically, we assume to work with a JJ in which the electrode $S_1$ is kept at a fixed temperature $T_1$, while $S_2$ has a floating temperature $T_2$. This can be accomplished by optimizing the volumes of the electrodes.
For the sake of readability, hereafter we will use an abbreviated notation in which $x$ and $t$ dependences are left implicit, namely, $T_2=T_2(x,t)$, $\varphi=\varphi(x,t)$, and $V=V(x,t)$. 
A characteristic length scale for the thermalization in the diffusive regime can be estimated as the inelastic scattering length $\ell_{in}=\sqrt{\mathcal{D}\tau_s}$, where $\mathcal{D}=\sigma_N/(e^2N_F)$ is the diffusion constant (with $\sigma_N$ and $N_F$ being the electrical conductivity in the normal state and the density of states at the Fermi energy, respectively) and $\tau_s$ is the quasiparticle scattering lifetime~\cite{Kap76}. In Ref.~\cite{Gua18}, the value $\ell_{in}\simeq0.3\;\mu\text{m}$ was estimated for a Nb lead at $4.2\;\text{K}$. So, when only the length of the junction is much longer than this length scale, i.e., $L\gg\ell_{in}$, the electrode $S_2$ can be modelled as a one-dimensional diffusive superconductor at a temperature varying along $x$ direction~\cite{Gua18}. In this case the evolution of the temperature $T_2$ is given by the time-dependent diffusion equation~\cite{Gua18}
\begin{equation}
\frac{\mathrm{d} }{\mathrm{d} x}\left [\kappa( T_2 ) \frac{\mathrm{d} T_2}{\mathrm{d} x} \right ]+\mathcal{P}_{\text{in}}\left ( T_1,T_2,\varphi,V \right )-\mathcal{P}_{e\text{-ph}}\left ( T_2,T_{\text{bath}}\right )=c_v(T_2)\frac{\mathrm{d} T_2}{\mathrm{d} t}.
\label{ThermalBalanceEq}
\end{equation}
Here, the rhs represents the variations of internal energy density of $S_2$ and the lhs terms indicate the spatial heat diffusion, taking into account the inhomogeneous electronic heat conductivity, $\kappa(T_2)$, and both the phase-dependent incoming, i.e., $\mathcal{P}_{\text{in}}\left ( T_1,T_2,\varphi,V \right )$, and the outgoing, i.e., $\mathcal{P}_{e\text{-ph}}\left ( T_2,T_{\text{bath}}\right )$, thermal power densities in $S_2$. 
The phase-dependent thermal power density flowing from $S_1$ to $S_2$ reads
\begin{equation}\label{Pt}
\mathcal{P}_{\text{in}}( T_1,T_2,\varphi,V)=\mathcal{P}_{\text{qp}}( T_1,T_2,V)-\cos\varphi \;\mathcal{P}_{\cos}( T_1,T_2,V).
\end{equation}
In the adiabatic regime~\cite{Gol13}, that is when the voltage drop is smaller than the relevant energy scales in the system, $eV\ll \text{min} \left \{ k_BT_1, k_BT_2, \Delta_1(T_1), \Delta_2(T_2) \right \}$, the contributions $\mathcal{P}_{\text{qp}}$ and $\mathcal{P}_{\cos}$ can be written as
\begin{equation}\label{Pqp}
\mathcal{P}_{\text{qp}}(T_1,T_2,V)\!=\!\frac{1}{e^2R_ad_2}\!\int_{-\infty}^{\infty} \!\!\! d\varepsilon\mathcal{N}_1 ( \varepsilon-eV ,T_1 )\mathcal{N}_2 ( \varepsilon ,T_2 )(\varepsilon-eV) [ f ( \varepsilon-eV ,T_1 ) -f ( \varepsilon ,T_2 ) ] ,
\end{equation}
\begin{eqnarray}\label{Pcos}\nonumber
\mathcal{P}_{\text{cos}}( T_1,T_2,V )=&&\frac{1}{e^2R_ad_2} \int_{-\infty}^{\infty}d\varepsilon \mathcal{N}_1 ( \varepsilon-eV ,T_1 )\mathcal{N}_2 ( \varepsilon ,T_2 )\\
&&\times\frac{\Delta_1(T_1)\Delta_2(T_2)}{\varepsilon}[ f ( \varepsilon-eV ,T_1 ) -f ( \varepsilon ,T_2 ) ],
\end{eqnarray}
where $f ( E ,T )$ is the Fermi distribution function and $\mathcal{N}_j\left ( \varepsilon ,T \right )=\left | \text{Re}\left [ \frac{ \varepsilon +i\gamma_j}{\sqrt{(\varepsilon +i\gamma_j) ^2-\Delta _j\left ( T \right )^2}} \right ] \right |$ is the reduced superconducting density of state, with $\Delta_j\left ( T_j \right )$ and $\gamma_j$ being the BCS energy gap and the Dynes broadening parameter~\cite{Dyn78} of the $j$-th electrode, respectively~\footnote{We observe that the width $d_2$ of the electrode $S_2$ appears in Eqs.~\eqref{Pqp}-\eqref{Pcos} since we wrote the thermal balance equation, see Eq.~\eqref{ThermalBalanceEq}, in terms of volume power densities.}. 
Equation~\eqref{Pqp} describes heat power density carried by quasiparticles, namely, it is an incoherent flow of energy through the junction from the hot to the cold electrode~\cite{Mak65,Gia06}. Instead, Eq.~\eqref{Pcos} represents the phase-dependent part of heat transport originating from the energy-carrying tunneling processes involving recombination/destruction of Cooper pairs on both sides of the junction~\cite{Mak65,Gia06}. As we are going to discuss later more specifically, this term is responsible for the localized temperature modulation in the presence of a soliton. 

At this point we make a step backward to clarify better what we mean when we refer to heat transport due to a soliton. First of all, a soliton in a LJJ carries a magnetic flux quantum which is generated by a circulating supercurrent loop~\cite{Ust98}. These dissipationaless superconducting currents give no contribute in the thermal dynamics we are going to discuss. Furthermore, the energy transport in a thermal-biased JJ includes also a phase-dependent term due to energy-carrying tunneling processes involving directly Cooper pairs~\cite{Gol13}. However, since it is a purely reactive contribution~\cite{Vir17}, in writing the thermal balance equation, see Eq.~\eqref{ThermalBalanceEq}, we have to neglect it since it does not contribute to the average heat flux, which determines the stationary temperature profile $T_2$. So, when we mention ``soliton-induced'' thermal effects, we are still dealing with temperature variations produced by heat carried by quasiparticles flowing through the junction. This ``heat current'' is due to the imposed temperature gradient, i.e., the electrodes have to reside at different temperatures, but depends on the phase difference according to Eq.~\eqref{Pt}, as it was recently demonstrated in many caloritronics experiments~\cite{ForGia17}. This is why a soliton, which is nothing but a localized 2$\pi$ phase twist, can locally affects thermal transport and, therefore, the temperature of the junction. 

The term $\mathcal{P}_{e\text{-ph}}$ in Eq.~\eqref{ThermalBalanceEq} represents the energy exchange, per unit volume, between electrons and phonons in the superconductor and reads~\cite{Pek09}
\begin{eqnarray}\label{Qe-ph}\nonumber
\mathcal{P}_{e\text{-ph}}=&&\frac{-\Sigma}{96\zeta(5)k_B^5}\int_{-\infty }^{\infty}dEE\int_{-\infty }^{\infty}d\varepsilon \varepsilon^2\textup{sign}(\varepsilon)M_{_{E,E+\varepsilon}}\Bigg\{ \coth\left ( \frac{\varepsilon }{2k_BT_{\text{bath}}}\right ) \\
&&\times \Big [ \mathcal{F}(E,T_2)-\mathcal{F}(E+\varepsilon,T_2) \Big ]-\mathcal{F}(E,T_2)\mathcal{F}(E+\varepsilon,T_2)+1 \Bigg\},
\end{eqnarray}
where $\mathcal{F}\left ( \varepsilon ,T_2 \right )=\tanh\left ( \varepsilon/2 k_B T_2 \right )$, $M_{E,{E}'}=\mathcal{N}_i(E,T_2)\mathcal{N}_i({E}',T_2)\left [ 1-\Delta ^2(T_2)/(E{E}') \right ]$, $\Sigma$ is the electron-phonon coupling constant, and $\zeta$ is the Riemann zeta function. Here, we are assuming that the lattice phonons of the superconductor are very well thermalized with the substrate that resides at $T_{\text{bath}}$, thanks to the vanishing Kapitza resistance between thin metallic films and the substrate at low temperatures~\cite{Gia06}.

Finally, in Eq.~\eqref{ThermalBalanceEq}, $c_v(T)=T\frac{\mathrm{d} \mathcal{S}(T)}{\mathrm{d} T}$ is the volume-specific heat capacity, with $\mathcal{S}(T)$ being the electronic entropy density of $S_2$~\cite{Sol16}
\begin{equation}
\mathcal{S}(T)\!=\!-4k_BN_F\int_{0}^{\infty}\!d\varepsilon \mathcal{N}_2(\varepsilon,T)\left\{ \left [ 1-f(\varepsilon,T) \right ] \ln\left [ 1-f(\varepsilon,T) \right ]+f(\varepsilon,T) \ln f(\varepsilon,T)\right \},
\label{Entropy}
\end{equation}
and $\kappa(T_2)$ is the electronic heat conductivity~\cite{For17}
\begin{equation}\label{electronicheatconductivity}
\kappa(T_2)=\frac{\sigma_N}{2e^2k_BT_2^2}\int_{-\infty}^{\infty}\mathrm{d}\varepsilon\varepsilon^2\frac{\cos^2\left \{ \text{Im} \left [\text{arctanh} \left (\frac{\Delta(T_2)}{\varepsilon+i\gamma_2} \right )\right ] \right \}}{\cosh ^2 \left (\frac{\varepsilon}{2k_BT_2} \right )}.
\end{equation}

For gaining insight in thermal transport through the junction, it only remains to include in Eq.~\eqref{ThermalBalanceEq} the specific phase $\varphi(x,t)$ for a LJJ given by solving numerically Eqs.~\eqref{SGeq} and~\eqref{bcSGeq}, with initial conditions $\varphi(\widetilde{x},0)=d\varphi(\widetilde{x},0)/d\widetilde{t}=0\quad \forall \widetilde{x}\in[0-\widetilde{L}]$. 
%
%

\section{Numerical results}
\label{Results}\vskip-0.2cm 
%

In the present study, we consider an Nb/AlO$_x$/Nb tunnel LJJ characterized by a normal resistance per area $R_a=50~\Omega~\mu\text{m}^2$ and a specific capacitance $C=50~fF/\mu \text{m}^2$. The linear dimensions $(L\times W\times d_2)$ of the junction are set to $(150\times0.5\times0.1)\;\mu\text{m}$ and $d=1\;\text{nm}$ is the thickness of the insulating layer. 

For Nb electrodes, we assume $\lambda_{L}^0=80\;\text{nm}$, $\sigma_N=6.7\times10^6 \;\Omega^{-1}\text{m}^{-1}$, $\Sigma=3\times10^9\;\textup{W}\textup{m}^{-3}\textup{ K}^{-5}$, $N_F=10^{47}\;\textup{ J}^{-1}\textup{ m}^{-3}$, $\Delta_1(0)=\Delta_2(0)=\Delta=1.764k_BT_c$ (with $T_c=9.2\;\text{K}$ being the Nb critical temperature), and $\gamma_1=\gamma_2=10^{-4}\Delta$.

We impose a thermal gradient across the system, specifically, the bath resides at $T_{\text{bath}}=4.2\;\text{K}$, and $S_1$ resides at a temperature $T_1=7\;\text{K}$ kept fixed throughout the computation. This value of the temperature $T_1$ assures the maximal soliton-induced heating in $S_2$, for a bath residing at $T_{\text{bath}}=4.2\;\text{K}$~\cite{Gua18}. Nonetheless, the local heating that we are going to discuss could be enhanced by lowering the bath temperature and then adjusting the temperature $T_1$. However, we stress that reducing the working temperatures could lead to a significantly longer thermal response time~\cite{Gua17}.

\begin{figure}[t!!]
\centering
\includegraphics[width=\columnwidth]{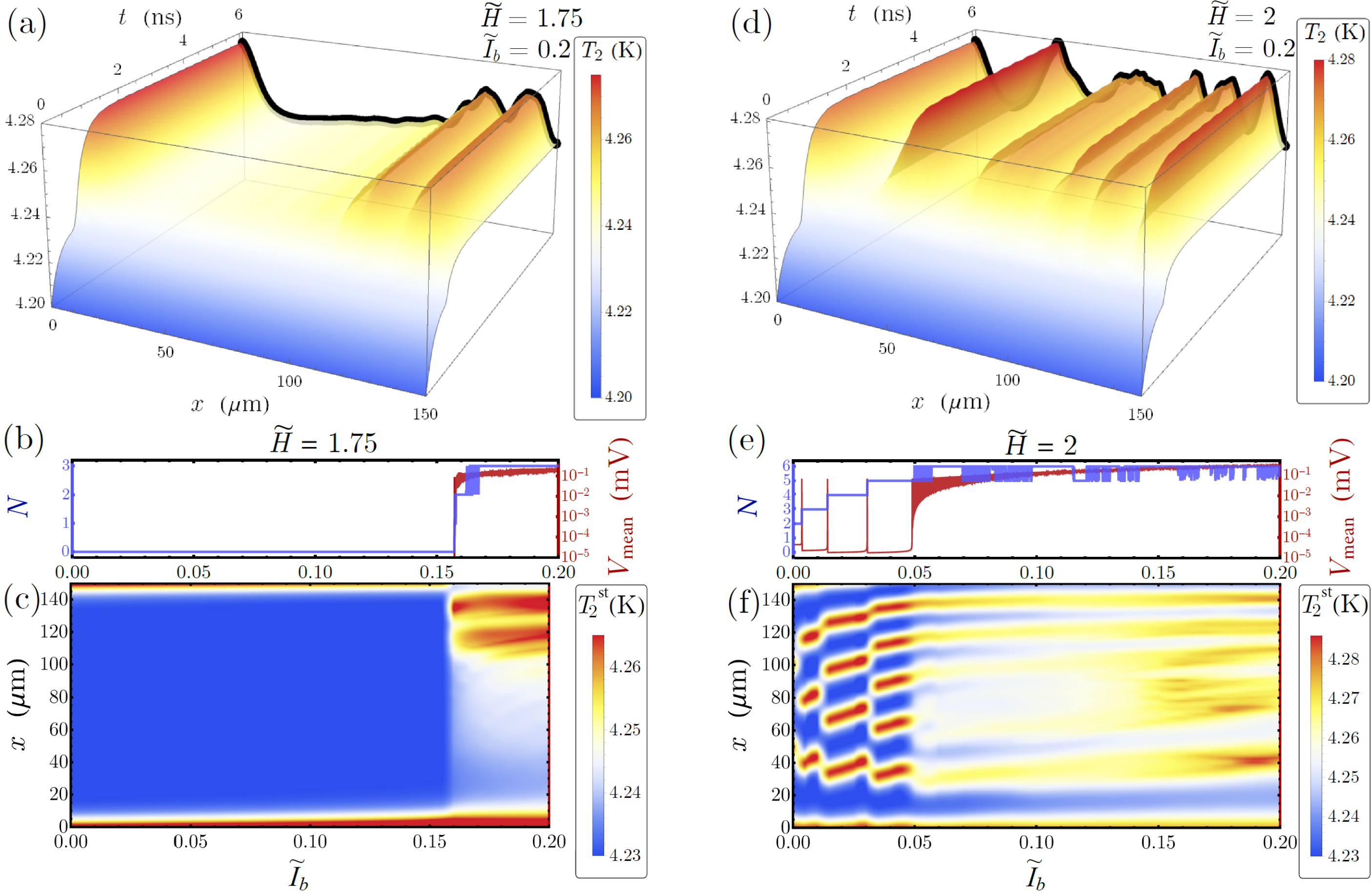}
\caption{(a) and (c), Temperature evolution for a fixed bias current $\widetilde{I}_b=0.2$, at $\widetilde{H}=1.75$ and $\widetilde{H}=2$, respectively. (b) and (e), Soliton number, $N$, and mean voltage drop, $V_{\text{mean}}$, as a function of the bias current $\widetilde{I}_b$ at $\widetilde{H}=1.75$ and $\widetilde{H}=2$, respectively. (c) and (f), Stationary temperature profile $T^{\text{st}}_2(x)$ as a function of $\widetilde{I}_b$, at $\widetilde{H}=1.75$ and $\widetilde{H}=2$, respectively.} 
\label{Figure02}
\end{figure}

The electronic temperature $T_2(x,t)$ of the electrode $S_2$ is the key quantity to master thermal transport across the junction, since it can modulate under the influence of both the external magnetic field and the bias current. By taking into account the specific temperature-dependence in both $t_d(T_1,T_2)$ and $J_c( T_1,T_2)$~\cite{Gia05,Bos16}, we can estimate the values of the parameters of the system: $\lambda_{_{\text{J}}}\simeq7.1\;\mu\text{m}$, $\omega_p\simeq1.3\;\text{THz}$, $H_{c,1}\simeq5.1\;\text{Oe}$, and $\alpha\simeq0.3$. According to this $\lambda_{_{\text{J}}}$ value, the length of the junction in normalized units reads $\widetilde{L}=L/\lambda_{_{\text{J}}}\simeq21$, while the value of the damping parameter $\alpha\simeq0.3$ gives an underdamped dynamics. Anyway, despite the temperature changes, during the time evolution these parameters are assumed constant, since they weakly depend on $T_2$ in the range of values that we are going to discuss.

The evolution of the temperature $T_2$ at fixed values of bias current $\widetilde{I}_b=0.2$ and magnetic field $\widetilde{H}=1.75$ is shown in Fig.~\ref{Figure02}(a). Here we are assuming to switch on the external magnetic field only when $T_2$ has reached a steady value $T_{2_s}$ between $T_{\text{bath}}$ and $T_1$. 
We observe that the temperature of $S_2$ locally increase at the left junction edge, $x=0$, and is double-peaked close to the right junction edge, $x=L$. This temperature distribution is a consequence of the soliton evolution triggered by the non-zero bias current. In fact, despite the magnetic field is under the threshold value, $\widetilde{H}<\widetilde{H}^{\text{thr}}$ (where $\widetilde{H}^{\text{thr}}=2$), the imposed bias current is high enough to induce a flux-flow regime. The bias-current conditions giving a flux-flow regime can be grasped by studying the number of solitons and the mean voltage drop across the junction. The number of solitons $N$ excited along the junction can be roughly evaluated through the quantity~\cite{Kup06}
\begin{equation}
N(t)=\left \lfloor \frac{\varphi(L,t)-\varphi(0,t)}{2\pi} \right\rfloor,
\label{SolitonNumber}
\end{equation}
where $\left \lfloor ... \right \rfloor$ stands for the integer part of the argument. 
The mean voltage across the junction can be estimated as
\begin{equation}
V_{\text{mean}}(t)=\frac{1}{L}\int_{0}^{L}\frac{\Phi_0}{2\pi}\frac{\mathrm{d} \varphi \left ( x,t \right )}{\mathrm{d} t}dx,
\label{meanvoltage}
\end{equation}
according to the \emph{a.c. Josephson relation}~\cite{Bar82}. 

In Fig.~\ref{Figure02}(b) we show the behaviour of both the number of solitons (left vertical scale, blue line) and the mean voltage drop (right vertical scale, red line) by quasi-adiabatically, i.e., very slowly, increasing the bias current, at a fixed external magnetic field $\widetilde{H}=1.75$. 
We observe that at a low bias current the system is in the \emph{Meissner state}~\cite{Kup06}, that is the fluxon-free state, corresponding to $N = 0$. 
Instead, when $\widetilde{I}_b\gtrsim0.156$ a flux-flow regime is established. In this case, we obtain $N>0$, that is solitons fill the junction moving from the left towards the right edge of the device driven by the current. In this regime a non-zero mean voltage drop appears, so that the larger the bias current, the higher the speed of soliton and therefore the larger $V_{\text{mean}}$. 

Then, despite the fast soliton dynamics, the flux-flow regime triggered by the bias current results in a peculiar temperature profile along the junction. In Fig.~\ref{Figure02}(a), we also highlight with a black solid curve the stationary temperature profile, $T^{\text{st}}_2(x)$, since we are going to show shortly how this stationary profile modifies as the bias current changes. In fact, Fig.~\ref{Figure02}(c) is drawn collecting several stationary temperature profiles $T^{\text{st}}_2(x)$ by changing the bias current, at $\widetilde{H}=1.75$.
\begin{figure}[t!!]
\centering
\includegraphics[width=0.6\columnwidth]{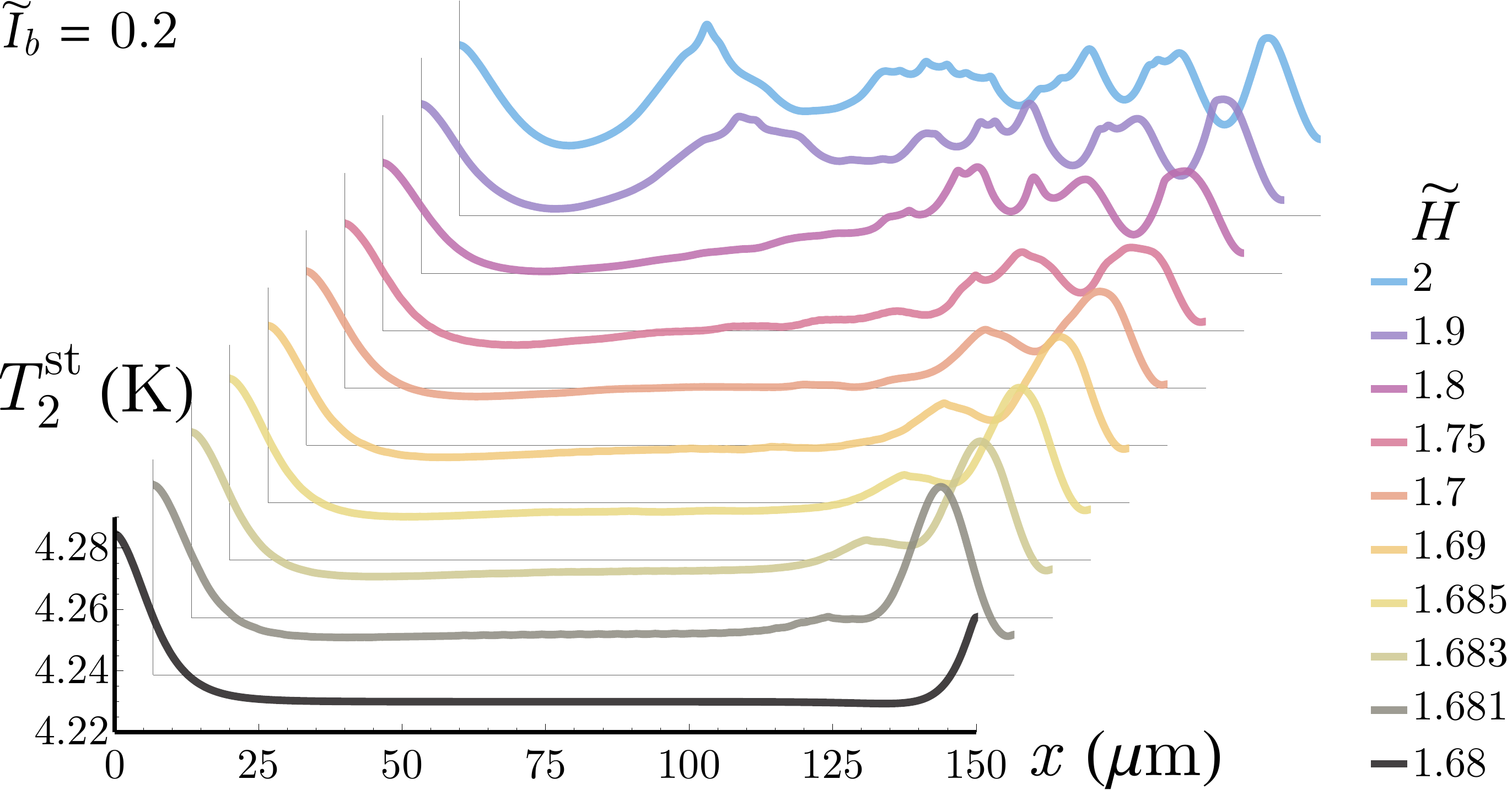}
\caption{Stationary temperature $T^{\text{st}}_2$ as a function of $x$, at different values of magnetic field and for a fixed $\widetilde{I}_b=0.2$.} 
\label{Figure03}
\end{figure}
%
In the flux-flow regime, that is for $\widetilde{I}_b\gtrsim0.156$, we observe that the average temperature along the junction increases and that two temperature peaks arise close to the right junction edge. This means that, once the flux-flow mode is triggered, the peculiar peaked behaviour of the temperature depends little on the bias current, at least in the range of values that we are taking into account. Conversely, the stationary temperature profile could be mostly dependent on the value set for the external magnetic field.
So, we explored also the thermal effects in the flux-flow regime as a slightly higher value of the magnetic field is set, specifically, we impose $\widetilde{H}=\widetilde{H}^{\text{thr}}$. 

In Fig.~\ref{Figure02}(d) we show the time evolution of the temperature profile for $\widetilde{I}_b=0.2$ and $\widetilde{H}=2$. 
Now the situation is somewhat different with respect to what we shown previously when we set an under-threshold magnetic field value. In fact, now when the magnetic field is switched on several temperature peaks come into being along the junction. The analysis of the number of solitons $N$ and the mean voltage peak $V_{\text{mean}}$ as a function of $\widetilde{I}_b$ shown in Fig.~\ref{Figure02}(e), reveals that there is no Meissner state in this case, since already for $\widetilde{I}_b=0$ two solitons populate the junction, i.e., $N=2$. Then, by slightly increasing the bias current, solitons are pushed rightwards becoming more tightly-packed. At a certain point, a new soliton enters from the left edge. Each time that a new soliton enters, the phase configuration abruptly changes and a peak in the mean voltage drop appears. Finally, for $\widetilde{I}_b\gtrsim0.049$ the flux-flow mode begins and a non-negligible mean voltage drop definitively appears. 
The stationary temperature profile, $T^{\text{st}}_2(x)$, as a function of the bias current evidences two different regimes, see Fig.~\ref{Figure02}(f). For $\widetilde{I}_b<0.049$, the configuration of solitons is stationary, so that the temperature rises just in correspondence to each soliton, as it was already observed in Ref.~\cite{GuaSolBraGia18} in the absence of a bias current. Instead, for $\widetilde{I}_b\gtrsim0.049$, the flux-flow regime starts, so that solitons rapidly move along the junction, but despite this, a multi-peaked temperature configuration still emerges, reminiscent of the multi-peaked structure before the flux-flow is triggered. 

The effect of the magnetic field on the temperature $T_2$ is well outlined in Fig.~\ref{Figure03}, where we show the stationary temperature profile $T^{\text{st}}_2(x)$, by changing the intensity of the underthreshold magnetic field, at a fixed $\widetilde{I}_b=0.2$. At $\widetilde{H}=1.68$ the temperature distribution is asymmetric, but there are no temperature peaks along the junction, since the system is in the Meissner state. Instead, by increasing further the magnetic field, the system goes into the flux-flow regime and some temperature peaks appear, so that the stronger the magnetic field, the greater the number of temperature peaks.

\begin{figure}[t!!]
\centering
\includegraphics[width=\columnwidth]{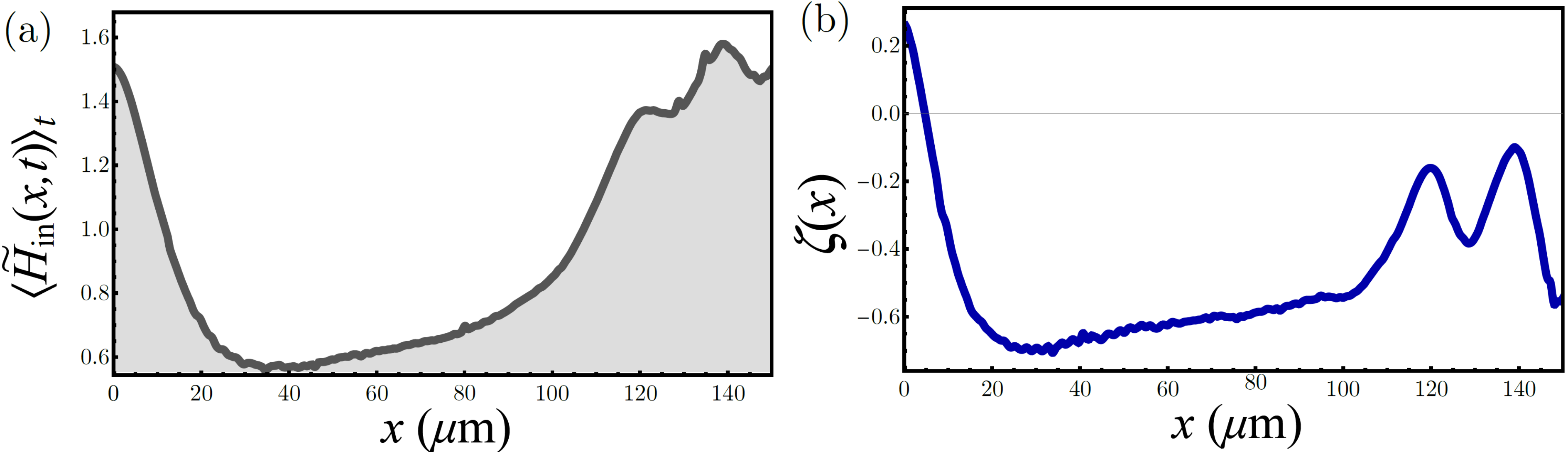}
\caption{Time averages of the normalized local magnetic field (a), see Eq.~\eqref{IntH}, and the cosine of the phase (b), see Eq.~\eqref{IntCos}, as a function of $x$, for $\widetilde{H}=1.75$ and $\widetilde{I}_b=0.16$. } 
\label{Figure04}
\end{figure}

Although the timescale of the solitonic evolution is generally shorter than the timescale of thermal relaxation processes~\footnote{The Swihart velocity can be of the order of $\bar{c}\sim10^7\;\text{m}/\text{s}$. Thus, a soliton moving at a speed equal, for instance, to $0.1\,\bar{c}$ takes approximatively $7\;\text{ps}$ to cover a distance roughly equal to the length scale of the system $\sim\lambda_J\simeq7\;\mu\text{m}$. Conversely, in a Nb-junction the estimated thermal response time is of the order of a fraction of nanosecond~\cite{GuaSol18}.}, the results discussed so far show that soliton-induced thermal effects emerge also in the flux-flow regime, that is when a chain of solitons rapidly moves along the junction. 
To understand the physical origin of this behaviour, we define two quantities giving information on both the soliton position and the distribution of the phase-dependent component of heat current flowing through the system. 
The soliton configurations are well depicted by the space derivative of the phase, $\frac{\partial \varphi(x,t)}{\partial x}$, since it is proportional to the local magnetic field according to the relation~\cite{Bar82}
\begin{equation}
H_{\text{in}}(x,t)=\frac{H_{c,1}\lambda_J}{2}\frac{\partial \varphi (x,t) }{\partial x}.
\label{localmagneticfield}
\end{equation}
Thus, to understand the thermal response in the flux-flow regime, we can define the time average of the normalized local magnetic field along the junction as
\begin{equation}
\left \langle \widetilde{H}_{\text{in}}(x,t) \right \rangle _t =\frac{1}{T_p}\int_{t_0}^{t_0+T_p}\frac{\partial \varphi (x,t) }{\partial x}dt,
\label{IntH}
\end{equation}
where $t_0$ is the time at which flux-flow starts and $T_p$ is a much longer time than the typical timescale of the solitonic evolution. The $\left \langle \widetilde{H}_{\text{in}}(x,t) \right \rangle _t$ profile obtained for $\widetilde{H}=1.75$ and $\widetilde{I}_b=0.16$ is shown in Fig.~\ref{Figure04}(a). We note that this function is peaked at $x=0$ and also that it significantly enhances close to the right junction edge. This picture confirms that, in the range of bias current and magnetic field under investigation, during the time evolution some solitons are preferentially localized in the right part of the junction. In fact, during the flux-flow evolution, for a magnetic field close to the critical value and for a low enough bias current, the dynamics we observed is composed by a soliton chain ``surfing'' on a standing solitons background~\footnote{A similar surfing soliton dynamics was also previously discussed investigating the behaviour of SG chiral soliton lattice in helimagnetic structures in the presence of a magnetic field~\cite{Kin15}.}. This standing solitons configuration is responsible for the peaked temperature profile close to the right junction edge discussed so far.

Since the peaked behaviour of the temperature profile along the junction can be mainly ascribed to the phase-dependent contribution in Eq.~\eqref{Pt}, namely, to the term $-\cos\varphi \;\mathcal{P}_{\cos}$, it can be useful to define the quantity
\begin{equation}
\zeta(x)=-\frac{1}{T_p}\int_{t_0}^{t_0+T_p}\cos \left [\varphi\left ( x,t \right ) \right ]dt,
\label{IntCos}
\end{equation}
which behaviour as a function of $x$, for $\widetilde{H}=1.75$ and $\widetilde{I}_b=0.16$, is shown in Fig.\ref{Figure04}(b). By comparing this curve with results in Figs.~\eqref{Figure02}(a) and (c), the response of the temperature $T_2$ reflects the behavior of the function $\zeta(x)$, which is also clearly double-peaked near the right edge of the junction.

\begin{figure}[t!!]
\centering
\includegraphics[width=0.6\columnwidth]{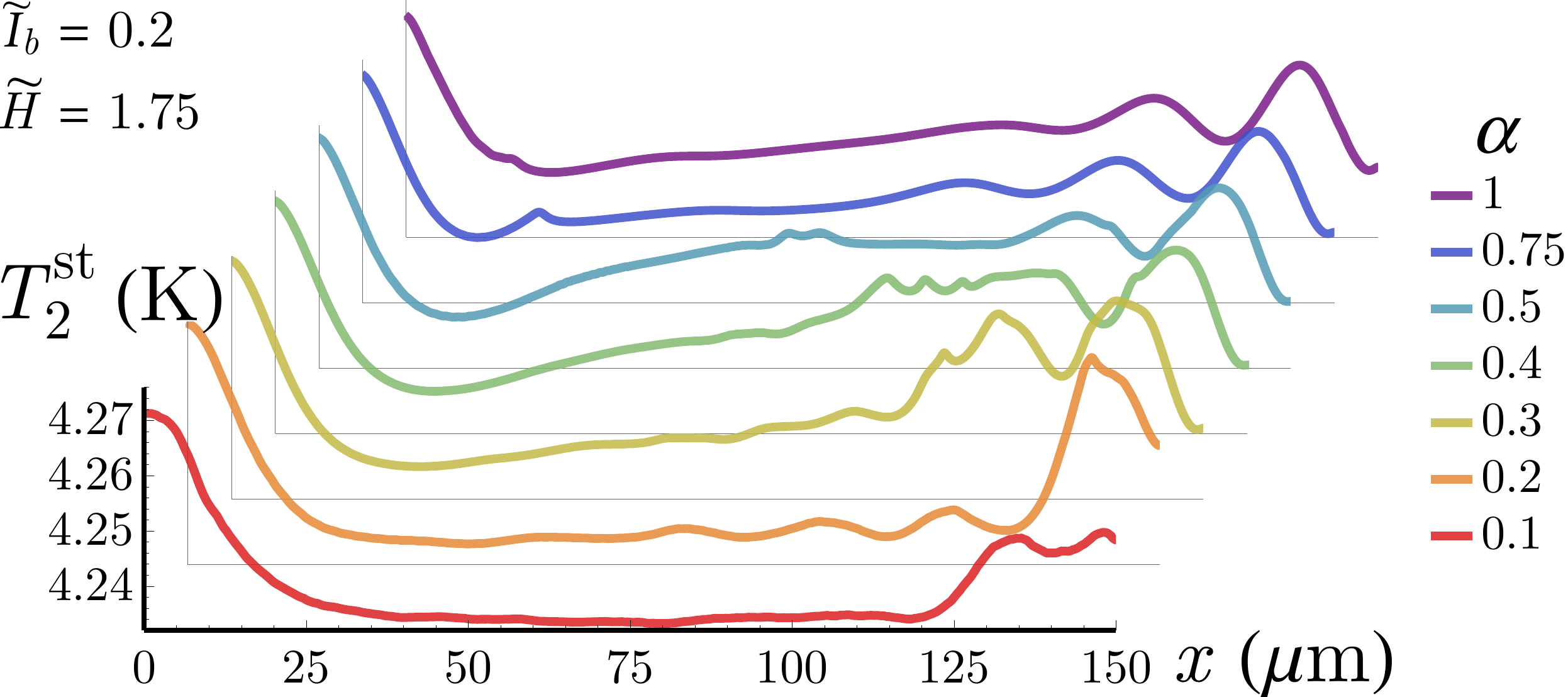}
\caption{Stationary temperature $T^{\text{st}}_2$ as a function of $x$, at fixed values of $\widetilde{I}_b=0.2$ and $\widetilde{H}=1.75$, by varying the damping parameter $\alpha\in[0.1-1]$.} 
\label{Figure05}
\end{figure}
Finally, since the temperature reached locally depends on dynamical aspects of the soliton evolution, we investigate how the damping affecting the phase dynamics influences thermal transport. As a friction parameter, $\alpha$ opposes the variations of $\varphi$. Then, we present in Fig.~\ref{Figure05} the stationary temperature profile as a function of $\alpha$, at fixed values of the bias current and the magnetic field, $\widetilde{I}_b=0.2$ and $\widetilde{H}=1.75$, respectively. We observe that in the range of $\alpha$ values taken into account, the overall thermal behaviour is only slightly modified by a change in the damping parameter. By increasing $\alpha$ the shape of the two temperature peaks becomes clearer, and the small ripples in the temperature profiles, probably due to leftward retro-reflected phase components, tend to fade. For higher damping values (i.e., $\alpha\gg1$ not showed in the figure) solitons slow so much down that timescales of soliton dynamics and thermal relaxation process become comparable, so that the temperature modulates in time following the moving solitons evolution and no stationary temperature profiles emerges in the flux-flow regime. 

Finally, we observe that in this work we are neglecting noisy effects eventually induced by stochastic thermal fluctuations affecting both the phase dynamics and the thermalization process. In other words, we could in principle include stochastic white-noise sources into both the SG equation~\cite{Fed07,Aug09,Gua13,Val14} and the thermal relaxation model Eq.~\eqref{ThermalBalanceEq} (e.g., Ref.~\cite{Bra18}). Nevertheless, according to the fluctuation-dissipation theorem, we stress that the amplitude of phase and temperature fluctuations can be reduced by increasing, respectively, the normal-state resistance of the junction~\cite{Bar82} and the heat capacitance (that is, the volume) of the thermally floating electrode~\cite{Bra18}. Anyway, we will present a more detailed stochastic analysis of thermal effects in a temperature-biased LJJ in a forthcoming paper.

\section{Conclusions}
\label{Conclusions}

In this paper we study thermal transport in a temperature-biased LJJ operating in the flux-flow regime. Specifically, the electrodes forming the junction reside at different temperatures and the device is driven by both a bias current, $\widetilde{I}_b$, and a magnetic field, $\widetilde{H}$. The magnetic field used is close to the threshold value, $\widetilde{H}^{\text{thr}}=2$, the latter being the magnetic field, in normalized units, above which solitons enter a junction also in the absence of a bias current. In this case the fluxon density along the junction is not very high and we can suppose to highlight soliton-induced well-localized thermal effects in the system. 

We observe that, by increasing the bias current, as soon as the junction enters in the flux-flow mode, the temperature profile along the junction modifies. Despite in this regime solitons rapidly move along the junction, the temperature tends to rise just in specific points, depending mainly on the value of the applied magnetic field. In fact, by increasing further the magnetic field a multi-peaked structure in the temperature emerges. Specifically, we compare thermal evolutions obtained by setting the normalized magnetic field to an underthreshold value $\widetilde{H}=1.75$ and to $\widetilde{H}=\widetilde{H}^{\text{thr}}=2$.
We study also the number of fluxons and the mean voltage drop across the junction as a function of the bias current, to highlight the $\widetilde{I}_b$ values triggering the flux-flow regime. 

Finally, we investigate how the friction affecting the phase dynamics enters into play in the thermal evolution of the system. Specifically, we study how a change in the damping parameter $\alpha$ influences the stationary temperature profiles along the junction. 

Our findings are important for understanding the interplay between soliton dynamics and thermal evolution in a LJJ, in those conditions in which, despite the extremely rapid time evolution of the Josephson phase should suggest the absence of relevant thermal effects, we are still able to highlight the emergence of intriguing phenomena. Markedly, we observe that, starting from a homogeneous system, we finally observe some inhomogeneously-distributed temperature profiles. The initial symmetry of the system is indeed broken by the applied bias current, which forces the soliton to move in a specific direction. Since thermal effects that we discussed are independent on the polarity of the magnetically-excited soliton, similar temperature configurations can be obtained by inverting both the bias current flowing direction and the polarity of the solitons, the latter depending on the direction of the in-plane external magnetic field.

\section{Acknowledgments}
\label{acknowledgments}

We thank G. Filatrella for the useful discussions. A.B. and F.G. acknowledge the European Research Council under the European Union's Seventh Framework Program (FP7/2007-2013)/ERC Grant agreement No.~615187-COMANCHE and the Tuscany Region under the PAR FAS 2007-2013, FAR-FAS 2014 call, project SCIADRO, for financial support.
A.B. acknowledges the CNR-CONICET cooperation programme ``Energy conversion in quantum nanoscale hybrid devices'' and the Royal Society though the International Exchanges between the UK and Italy (grant IES R3 170054).

\bibliographystyle{iopart-num}

\providecommand{\newblock}{}

\end{document}